\documentclass[12pt]{article}
\usepackage[english,german,french,polish]{babel}
\usepackage[T1]{fontenc}
\usepackage{amsfonts}

\begin{document}

\selectlanguage{english}

\baselineskip 0.73cm
\topmargin -0.4in
\oddsidemargin -0.1in

\let\ni=\noindent

\renewcommand{\thefootnote}{\fnsymbol{footnote}}

\newcommand{\SM}{Standard Model }

\pagestyle {plain}

\setcounter{page}{1}



~~~~~~
\pagestyle{empty}

\begin{flushright}
IFT-- 08/9
\end{flushright}

\vspace{0.4cm}

{\large\centerline{\bf Again on hidden sector of the Universe, accessible}}

{\large\centerline{\bf through photonic portal}}

\vspace{0.5cm}

{\centerline {\sc Wojciech Kr\'{o}likowski}}

\vspace{0.3cm}

{\centerline {\it Institute of Theoretical Physics, University of Warsaw}}

{\centerline {\it Ho\.{z}a 69, 00--681 Warszawa, ~Poland}}

\vspace{0.6cm}

{\centerline{\bf Abstract}}

\vspace{0.2cm}

\begin{small}

\begin{quotation}

We describe more precisely  the mechanism of spontaneous mass generation in the previously proposed model
of hidden sector of the Universe. The hidden sector is conjectured to consist of sterile spin-1/2 fermions 
(sterinos) whose mass is spontaneously generated by sterile scalar bosons (sterons), while their interactions 
are mediated by sterile antisymmetric-tensor bosons ({\it A} bosons). Such a sector communicates with the 
familiar Standard Model sector through a photonic portal acting weakly between both sectors (but stronger 
than the universal gravity). This is due to photons which, beside sterinos and sterons, contribute to the source of 
the {\it A} bosons. Sterinos can be candidates for thermal cold dark matter. They can be also 
produced in sterino-antisterino pairs through virtual photons emitted in high-energy collisions of Standard 
Model charged particles. Sterinos display a tiny magnetic moment spontaneously generated by sterons.
 
\vspace{0.6cm}

\ni PACS numbers: 14.80.-j , 04.50.+h , 95.35.+d 

\vspace{0.3cm}

\ni September 2008

\end{quotation}
 
\end{small}

\vfill\eject

\pagestyle {plain}

\setcounter{page}{1}

\vspace{0.5cm}

\ni {\bf 1. Introduction}

\vspace{0.3cm} 

In a recent work [1], new {\it supplemented Maxwell equations} were proposed to unify the \SM electrodynamics (emerging after the electroweak symmetry is sponta\-neosly broken by the \SM Higgs mechanism) with a new sterile sector dyna\-mics. Such a sector (the {\it hidden sector}) is conjectured to be represented by sterile spin-1/2 
fermions ({\it sterinos}) whose mass is spontaneously generated by sterile scalar bosons ({\it sterons}), whereas their interactions are mediated by sterile antisymmetric-tensor bosons ({\it A bosons}). Denoting the corresponding sterile fields by $\psi, \varphi$ and $A_{\mu \nu}$, respectively, the proposed equations read

\begin{equation}
\partial_\nu (F^{\mu\,\nu} +  \sqrt{f} \varphi A^{\mu\,\nu}) = -j^\mu 
\end{equation}

\ni and

\begin{equation}
(\Box - M^2)A_{\mu\,\nu} = - \sqrt{f} (\varphi F_{\mu\,\nu} + \zeta \bar\psi \sigma_{\mu\,\nu} \psi)\;, 
\end{equation}

\ni where  $j_\mu$ and 

\begin{equation} 
F_{\mu\,\nu} = \partial_\mu A_\nu - \partial_\nu A_\mu 
\end{equation}

\ni are the \SM electric current and electromagnetic field, while $M$ denotes a large mass scale and $f>0$ as well as $\zeta$ stand for unknown dimensionless constants. Finally, $\sigma_{\mu\,\nu}  = (i/2)[\gamma_\mu,\gamma_\nu]$ with $(\gamma^\mu) = (\beta,\beta \vec{\alpha})$. The constant $f/4\pi$ is expected to be small (but large enough to make new interactions stronger than the universal gravity). For the steron field it is assumed that

\begin{equation} 
\varphi \equiv <\!\!\varphi\!\!>_{\rm vac} + \varphi^{\rm (ph)}\,,
\end{equation}

\ni where $<\!\!\varphi_{\rm vac}\!\!> \neq 0$ is a spontaneously nonzero vacuum expectation value of the field $\varphi$ and the physical part of $\varphi$ is denoted by $\varphi^{\rm (ph)}$. Then, the massdimensional constant $<\!\!\varphi_{\rm vac}\!\!> \neq 0$ generates spontaneously the sterino mass.

The quanta of sterile mediating field $A_{\mu \nu}$ (having dimension one) or {\it A} bosons get spin 1 and parity + or -- , depending whether they correspond to the (three-dimensional) vector or axial field, $A_{k 0}(x)$ or $\varepsilon_{k l m} A^{l m}(x)$, respectively ($ k = 1,2,3$). The mass $M$ of these quanta is also spontaneously generated by $<\!\!\varphi_{\rm vac}\!\!> \neq 0$. 

If momentum transfers mediated through the field $A_{\mu \nu}$ in virtual states are negligible {\it versus} the mass scale $M$, then one gets from Eq. (2) that approximately 

\vspace{0.2cm}

\begin{equation} 
A_{\mu\,\nu} \simeq \frac{\sqrt{f}}{M^2}  (\varphi F_{\mu\,\nu} + \zeta \bar\psi \sigma_{\mu\,\nu} \psi)\;, 
\end{equation}

\ni what reduces Eq. (1) to the form

\begin{equation} 
 \partial_\nu \left[F^{\mu \nu} + \frac{f}{M^2} \varphi (\varphi F^{\mu\,\nu} + \zeta \bar\psi \sigma^{\mu\,\nu} \psi)\right] \simeq - j^\mu 
\end{equation}

\vspace{0.2cm}

\ni or $\partial_\nu F^{\mu \nu} = -j^\mu - \delta j^\mu$, where the magnetic-like correction 

\begin{equation}
\delta j^\mu \equiv \sqrt{f} \partial_\nu(\varphi A^{\mu \nu}) \simeq 
\frac{f}{M^2}\,\partial_\nu \left[\varphi (\varphi F^{\mu \nu} + \bar\psi \sigma^{\mu\,\nu} \psi )\right]
\end{equation}

\vspace{0.2cm}

\ni to the \SM electric current $j^\mu$ appears as an effect of interaction with the sterile sector. Note that 
$\partial_\nu\delta j^\mu \equiv 0$ identically, while $\partial_\nu j^\mu = 0$ dynamically. The part of $\delta j^\mu $ (Eq.(7)) linear in $<\!\!\varphi\!\!>_{\rm vac}$ and independent of $\varphi^{(\rm ph)}$ is the sterino magnetic correction having the form $(f \zeta\!\! <\!\!\varphi\!\!>_{\rm vac}/M^2) \partial_\nu (\bar\psi \sigma^{\mu\,\nu} \psi)$. In the formal limit of $f/M^2\rightarrow 0$, Eq. (6) becomes the \SM Maxwell equation $\partial_\nu F^{\mu \nu} = -j^\mu$ with $ F_{\mu\,\nu} = \partial_\mu A_\nu - \partial_\nu A_\mu $.

We can see from Eqs. (1) and (2) that in our model the electromagnetic field $F^{\mu \nu}$ provides a link between the Standard-Model and sterile sectors of the Universe (the {\it photonic portal}). Its strength is proportional to $f/M^2$ and so, it is expected to be weak at low energies (but stronger than the universal gravity). For models postulating the alternative option of {\it Higgs portal} consult Ref. [2].

In the present paper, we will describe more precisely the mechanism of spontaneous mass generation for sterile particles by the nonzero vacuum expectation value $<\!\!\varphi\!\!>_{\rm vac} \neq 0$. Such a mass generation is conjectured in our model and formally introduced in Section 2. In Section 3, a new phenomenon of additional, finite renormalization appearing in our model is discussed, while in Section 4, sterinos as possible candidates for the thermal cold dark matter are presented. At the end of this Section there is a brief conclusion.

\vfill\eject

\vspace{0.3cm}

\ni {\bf 2. Spontaneous mass generation by $<\!\!\varphi\!\!>_{\rm vac} \neq 0$} 

\vspace{0.3cm}

Let us start with the following proposal of the unification of \SM electromagnetic Lagrangian $-(1/4)F_{\mu\,\nu} F^{\mu\,\nu} - j_\mu A^\mu$ (emerging after the spontaneous breakdown of the electroweak symmetry) with a new dynamics of our sterile fields $\psi, \varphi$ and $A_{\mu \nu}$:

\begin{eqnarray}
{\cal{ L }} & = &  -\frac{1}{4}F_{\mu\,\nu} F^{\mu\,\nu} - j_\mu A^\mu
- \frac{1}{2} \sqrt{f}\left(\varphi F_{\mu \nu} + \zeta \bar\psi \sigma_{\mu\,\nu} \psi \right) A^{\mu \nu}
\nonumber \\ & &   - \frac{1}{4} \left[\left(\partial_\lambda A_{\mu \nu}\right)\left(\partial^\lambda A^{\mu \nu}\right)  
- \frac{1}{2}\eta \varphi^2 A_{\mu \nu}A^{\mu \nu}\right]  \nonumber \\
 & & + \,\bar\psi \left(i \gamma^\lambda \partial_\lambda -  \xi \varphi\right) \psi + \left[\frac{1}{2}(\partial_\lambda\varphi) (\partial^\lambda\varphi ) - V(\varphi) \right] \;, 
\end{eqnarray}

\ni where $\varphi \equiv <\!\!\varphi\!\!>_{\rm vac} + \varphi^{(\rm ph)}$ and  $V(\varphi) \equiv -(1/2) \mu^2 \varphi^2 + (1/4) \lambda \varphi^4 $ with

\begin{equation}
\left(\frac{dV}{d\varphi}\right)_{\varphi = <\!\varphi\!>_{\rm vac}} \equiv <\!\!\varphi\!\!>_{\rm vac}\left(-\mu^2 + \lambda <\!\!\varphi\!\!>_{\rm vac}^2\right) = 0 
\end{equation}

\ni (in the tree approximation), giving

\begin{equation}
<\!\!\varphi\!\!>_{\rm vac} = \frac{\mu}{\sqrt \lambda} \,.
\end{equation}

\ni Here, $\mu$ and $\lambda > 0,\,\xi,\,\eta$ are unknown constants, massdimensional  and dimensionless, respectively. The corresponding total Lagrangian for the system consisting of both the Standard-Model and sterile sectors is the sum of Lagrangian (8) and the familiar \SM Lagrangian [3] with its electromagnetic part $-(1/4)F_{\mu\,\nu} F^{\mu\,\nu} - j_\mu A^\mu$ shifted in our case to the Lagrangian (8).

Thus, in the fifth and sixth term of  Lagrangian (8) we obtain

\begin{equation}
\xi \varphi = m_\psi + \xi \varphi^{(\rm ph)}
\end{equation}

\ni and

\begin{equation}
V(\varphi) = \frac{1}{2} m^2_\varphi\, \varphi^{({\rm ph})\,2} + \sqrt{\frac{\lambda}{2}}\, m_\varphi\, \varphi^{({\rm ph})\,3} + \frac{1}{4} \lambda \,\varphi^{({\rm ph})\,4} - \frac{m_\varphi^4}{16\lambda}\,,
\end{equation}

\ni where 

\begin{equation}
m_\psi = \xi <\!\!\varphi\!\!>_{\rm vac}\;,\;\, m_\varphi = \sqrt{2\lambda}<\!\!\varphi\!\!>_{\rm vac}
\end{equation}

\ni are the sterino and steron mass, respectively, spontaneously generated by $<\!\!\varphi\!\!>_{\rm vac} \neq 0$. Similarly, in the third term of Lagrangian (8) we have $\varphi^2 \equiv <\!\!\varphi\!\!>_{\rm vac}^2 + 2<\!\!\varphi\!\!>_{\rm vac} \varphi^{(\rm ph)} + \varphi^{(\rm ph)\,2}$ and so,

\begin{equation} 
\frac{1}{2} \eta \,\varphi^2 = M^2 + \sqrt{2\eta} \,M\, \varphi^{(\rm ph)} + \frac{1}{2} \eta \,\varphi^{(\rm ph)\,2}\,,  
\end{equation}

\ni where 

\begin{equation} 
M = \sqrt{\frac{\eta}{2}}<\!\!\varphi\!\!>_{\rm vac} 
\end{equation}

\ni is the mass of quanta of the sterile mediating field $A_{\mu \nu}$ or $A$ bosons, spontaneously generated by 
${<\!\!\varphi\!\!>_{\rm vac}} \neq 0$. 

With the use of spontaneously generated mass values $m_\psi,\,m_\varphi$ and $M$ we can rewrite our Lagrangian (8) as follows:

\begin{eqnarray}
\!\!{\cal{ L }}\!\! & \!\!=\!\! & \!\!-\frac{1}{4}F_{\mu\,\nu} F^{\mu\,\nu} - j_\mu A^\mu
- \frac{1}{2} \sqrt{f}\left(\varphi F_{\mu \nu} + \zeta \bar{\psi} \sigma_{\mu\,\nu} \psi \right) A^{\mu \nu}
\nonumber \\ \!\!\! & & \!\!\!\!-\frac{1}{4}\!\left[\!\left(\partial_\lambda A_{\mu \nu} \right)\!\left( \partial^\lambda A^{\mu \nu} \right)  
- \!(\! M^2 \!+\! \sqrt{2\eta}\, M \varphi^{(\rm ph)} \!+\! \frac{1}{2} \eta \varphi^{(\rm ph)\,2}) A_{\mu \nu}
A^{\mu \nu} \right] \!+\! \bar\psi(i \gamma^\lambda \partial_\lambda \!-\! m_\psi \!-\! \xi \varphi^{(\rm ph)}) \psi 
\nonumber \\ \!\!\!  & &\!\!\!+ \frac{1}{2}\left[(\partial_\lambda\varphi^{(\rm ph)}) (\partial^\lambda\varphi^{(\rm ph)}) - m^2_\varphi \,\varphi^{(\rm ph)\,2} - \sqrt{2\lambda}\, m_\varphi \,\varphi^{(\rm ph)\,3} \!-\!\frac{1}{2}\lambda\varphi^{(\rm ph)\,4}+ \frac{m^4_\varphi}{8 \lambda}\right] \;, \nonumber \\ & &
\end{eqnarray}

\ni where $\varphi \equiv {<\!\!\varphi\!\!>_{\rm vac}} + \varphi^{(\rm ph)}$, while $j_\mu$ denotes the \SM electric current.

Note that in the original Lagrangian (8), the scale invariance is broken solely by the nonzero massdimensional constant $\mu = \sqrt{\lambda}\,{<\!\!\varphi\!\!>_{\rm vac}} $ appearing in the potential $V(\varphi) \equiv -(1/2) \mu \varphi^2 + (1/4) \lambda \,\varphi^4$, and being spontaneously generated by the vacuum expectation value ${<\!\!\varphi\!\!>_{\rm vac}} \neq 0$. Thus, ${<\!\!\varphi\!\!>_{\rm vac}} \neq 0$ breaks spontaneously the scale invariance of our model, leading to the scale-violating Lagrangian (16) which contains three masses $m_\psi,\,m_\varphi$ and $M$ (all spontaneously generated by ${<\!\!\varphi\!\!>_{\rm vac}} \neq 0$).

The Lagrangian (16) implies the following four coupled field equations for $F_{\mu \nu},A_{\mu \nu},\psi $ and $\varphi^{(\rm ph)}$:

\vspace{-0.5cm}

\begin{equation} 
\partial_\nu(F^{\mu \nu} + \sqrt{f} \varphi A^{\mu \nu}) = - j^\mu \,,
\end{equation}

\begin{equation} 
\left[\Box - (\! M^2 + \sqrt{2\eta\,} M \,\varphi^{(\rm ph)} \!+\! \frac{1}{2} \eta \,\varphi^{(\rm ph)\,2})\right] A_{\mu \nu} = - \sqrt{f}(\varphi F_{\mu \nu} \!+\! \zeta \bar{\psi} \sigma_{\mu\,\nu} \psi) \,,
\end{equation}

\begin{equation} 
\left(i \gamma^\lambda \partial_\lambda \!-\! m_\psi \!-\! \xi \varphi^{(\rm ph)} \!-\! \frac{1}{2}\sqrt{f} \zeta
\sigma_{\mu \nu} A^{\mu \nu}\right)\psi = 0 \,,
\end{equation}

\begin{equation} 
\!\left(\!\Box \!-\! m^2_\varphi \!+\! { \frac{1}{4}} \eta A_{\mu \nu}A^{\mu \nu}\,\right)\!\varphi^{(\rm ph)} \!-\! 3\sqrt{\frac{\lambda}{2}} m_\varphi \varphi^{(\rm ph)\,2} \!-\! \lambda\varphi^{(\rm ph)\,3} \!=\! \frac{1}{2}\sqrt{f}\, F_{\mu \nu}A^{\mu \nu} \!-\! \sqrt{ \frac{\eta}{8}}  M A_{\mu \nu}A^{\mu \nu} \!+\! \xi \bar{\psi}\psi
\end{equation}

\ni with $\varphi \equiv {<\!\!\varphi\!\!>_{\rm vac}} + \varphi^{(\rm ph)}$.

In particular, consider the case of weak steron field $\varphi^{(\rm ph)}$, where --- in the steron functional representation $| \varphi^{\!(\rm ph)}\!\!>$ --- the ratio  $|\varphi^{(\rm ph)}(x)/{<\!\!\varphi\!\!>_{\rm vac}}|$ is small. This corresponds to processes with few non-relativistic physical sterons. Then, in the Lagrangian (16) , the additive parts following the expressions $(1/4) M^2 A_{\mu \nu}A^{\mu \nu} , -m_\psi \bar{\psi}\psi$ and $-(1/2)m^2_\varphi \varphi^{(\rm ph)\,2}$ in the $A_{\mu \nu}- , \psi- $ and $\varphi-$terms, respectively, can be  neglected{\footnote{In Ref. [1] these corrections are suppressed}}, what leads to the approximate Lagrangian

\begin{eqnarray}
{\cal{ L }} & = & -\frac{1}{4}F_{\mu\,\nu} F^{\mu\,\nu} - j_\mu A^\mu
- \frac{1}{2} \sqrt{f}\left(\varphi F_{\mu \nu} + \zeta \bar{\psi} \sigma_{\mu\,\nu} \psi \right) A^{\mu \nu}
\nonumber \\ & & -\frac{1}{4}\left[\left(\partial_\lambda A_{\mu \nu} \right)\left( \partial^\lambda A^{\mu \nu} \right) - M^2 A_{\mu \nu}A^{\mu \nu} \right] + \bar\psi(i \gamma^\lambda \partial_\lambda - m_\psi ) \psi \nonumber \\ & &+ \frac{1}{2}\left[(\partial_\lambda\varphi^{(\rm ph)}) (\partial^\lambda\varphi^{(\rm ph)}) - m^2_\varphi\, \varphi^{(\rm ph)\,2} + \frac{m^4_\varphi}{8 \lambda}\right]  
\end{eqnarray}

\ni with $\varphi \equiv {<\!\!\varphi\!\!>_{\rm vac}}+ \varphi^{(\rm ph)} $. In fact, these additive parts can be written down in a compact way as $(1/4) M^2 \,A_{\mu \nu}A^{\mu \nu}[(1+\varphi^{(\rm ph)}\!/\!\!\!<\!\!\varphi\!\!>_{\rm vac})^2 - 1] \simeq 0$ , $-m_\psi \bar{\psi}\psi \,\varphi^{(\rm ph)}\!/\!\!\!<\!\!\varphi\!\!>_{\rm vac} \simeq 0$ and $ -(1/2) m^2_\varphi \,\varphi^{(\rm ph)\,2}[(1+\varphi^{(\rm ph)}\!\!/2\!\!<\!\!\varphi\!\!>_{\rm vac})^2 - 1] + m^4_\varphi/16\lambda \simeq  m^4_\varphi/16\lambda$, respectively (the additive constant $ m^4_\varphi/16\lambda$ can be omitted in the Lagrangian). 

The simplified Lagrangian (21) gives the following four approximate field equations, coupled with each other:

\begin{equation} 
\partial_\nu(F^{\mu \nu} + \sqrt{f} \,\varphi\, A^{\mu \nu}) = - j^\mu \,,
\end{equation}

\begin{equation} 
\left(\Box - M^2\right) A_{\mu \nu} = \sqrt{f}(\varphi F_{\mu \nu} + \zeta\, \bar{\psi}\, \sigma_{\mu\,\nu} \psi) \,,
\end{equation}

\begin{equation} 
\left(i \gamma^\lambda \partial_\lambda - m_\psi -\frac{1}{2}\sqrt{f} \,\zeta\,
\sigma_{\mu \nu} A^{\mu \nu}\right)\psi = 0 \,,
\end{equation}

\begin{equation} 
\left(\Box - m^2_\varphi \right)\varphi^{(\rm ph)} = \frac{1}{2}\sqrt{f}\,F_{\mu \nu}A^{\mu \nu} \,,
\end{equation}

\ni where $\varphi \equiv {<\!\!\varphi\!\!>_{\rm vac}} + \varphi^{(\rm ph)}$. Notice that Eq. (22) is identical with Eq. (1) as well as with Eq. (17), and Eq. (23) identical with Eq. (2), but simpler than Eq. (18). Similarly, Eqs. (24) and (25) are simpler than Eqs. (19) and (20), respectively.

Eliminating $A^{\mu \nu}$ from the interaction term $-(1/2)\sqrt{f}(\varphi F_{\mu \nu} +\zeta\,\bar{\psi}\sigma_{\mu \nu}\psi) A^{\mu \nu}$ in the Lagrangian (21) by means of Eq. (5) (if in virtual states $-\Box$ is negligible {\it versus} $M^2$) and dividing the quadratic result by 2, we get approximately the following effective interaction Lagrangian:

\begin{equation} 
-\frac{1}{4}\frac{f}{M^2} (\varphi F_{\mu \nu} + \zeta\,\bar{\psi}\sigma_{\mu \nu}\psi)(\varphi F^{\mu \nu} + \zeta\,\bar{\psi}\sigma^{\mu \nu}\psi) 
\end{equation}

\ni with $\varphi \equiv {<\!\!\varphi\!\!>_{\rm vac}} + \varphi^{(\rm ph)}$. Its part linear in ${<\!\!\varphi\!\!>_{\rm vac}}$ and independent of $\varphi^{(\rm ph)}$ is the sterino magnetic interaction of the form

\begin{equation}
-\frac{f \zeta\!<\!\!\varphi\!\!>_{\rm vac}}{2 M^2}\,\bar{\psi}\sigma_{\mu \nu}\psi F^{\mu \nu} \equiv -\mu_\psi \,\bar{\psi}\sigma_{\mu \nu}\psi F^{\mu \nu} \,,
\end{equation}

\ni where 

\begin{equation}
\mu_\psi \equiv \frac{f \zeta\!<\!\!\varphi\!\!>_{\rm vac} }{2 M^2} = \frac{f\,\zeta\, \xi}{\eta m_\psi}
\end{equation}

\ni (see Eqs. (13) and (15)) is the sterino magnetic moment {\it generated spontaneously} by ${<\!\!\varphi\!\!>_{\rm vac}} \neq 0 $. If it happens that (1 to $10^3)m^2_\psi \sim M^2 \sim \,<\!\!\varphi\!\!>_{\rm vac}^2$ ({\it i.e.}, (1 to $10^3)\xi^2 \sim \eta/2 \sim 1$, see Eqs. (13) and (15)), then from Eq. (28) $\mu_\psi \sim$ (1 to $10^{-3/2})f \zeta/2m_\psi $. So, (1 to $10^{-3/2}) f \zeta $ might be thought to play the role of an electric charge carried by sterinos. Of course, it does not contribute to any electric current. In contrast, the sterino magnetic interaction is a real effect that can lead to the polarization of a sterino gas in external magnetic fields and so, its magnetization in these fields.

\vspace{0.3cm}

\ni {\bf 3. Primordial renormalization} 

\vspace{0.3cm}

When the interaction term in the Lagrangian (21) is replaced by its effective form (26) (and the 
$A_{\mu \nu}$-term omitted), we obtain the effective Lagrangian

\begin{eqnarray} 
{\cal{L}}_{\rm eff} & = &  -\frac{1}{4}F_{\mu \nu} F^{\mu \nu}- j_\mu A^{\mu} - \frac{1}{4}\frac{f}{M^2} (\varphi F_{\mu \nu} + \zeta\,\bar{\psi}\sigma_{\mu \nu}\psi)(\varphi F^{\mu \nu} + \zeta\,\bar{\psi}\sigma^{\mu \nu}\psi) \nonumber \\ & & + (\psi\!{\rm  -}\;{\rm and} \;\varphi^{(\rm ph)}\!{\rm -terms\; as\; in \;Eq.\; (21))}
\end{eqnarray}

\vspace{0.2cm}

\ni with $\varphi \equiv {<\!\!\varphi\!\!>_{\rm vac}} + \varphi^{(\rm ph)}$. Such an effective Lagrangian leads to a multiplicative constant factor at the electromagnetic kinetic term $-(1/4)F_{\mu \nu} F^{\mu \nu}$, because Eq. (29) can be rewritten as 

\begin{eqnarray} 
{\cal{L}}_{\rm eff} & = &  -\frac{1}{4}Z^{-1}F_{\mu \nu} F^{\mu \nu}- j_\mu A^{\mu} - \frac{1}{4}\frac{f}{M^2} (2\! <\!\!\varphi\!\!>_{\rm vac}\varphi^{(\rm ph)} + \varphi^{(\rm ph)\,2}) F_{\mu \nu} F^{\mu \nu} \nonumber \\ & &
- \frac{1}{2}\frac{f \zeta}{M^2} \varphi F_{\mu \nu} (\bar{\psi}\sigma^{\mu \nu}\psi)  - \frac{1}{4}\frac{f \zeta^2}{M^2} (\bar{\psi}\sigma_{\mu \nu}\psi)(\bar{\psi}\sigma^{\mu \nu}\psi) \nonumber \\ & & + (\psi\!{\rm -}\;{\rm and} \;\varphi^{(\rm ph)}\!{\rm -terms\; as\; in \;Eq.\; (21))}\,,
\end{eqnarray}

\vspace{0.2cm}

\ni where 

\begin{equation} 
Z^{-1} \equiv 1 + \frac{f}{M^2} <\!\!\varphi\!\!>^2_{\rm vac}\; > 1\,.
\end{equation}

\vspace{0.2cm}

\ni Note that $Z^{-1} = 1 + 2f\!/\eta$ due to Eq. (15). If it happens that $M^2 \sim  <\!\!\varphi\!\!>_{\rm vac}^2$ ({\it i.e.}, $\eta/2 \sim 1 $, see Eq. (15)), then $Z^{-1} \sim 1 + f$. 

It is possible to include this factor into the electromagnetic field $F_{\mu \nu}$ and the constants $e, f,$ and $\zeta$, performing the following finite (tree-order) renormalization which may be called {\it primordial renormalization} (or $P$-renormalization):

\begin{equation}  
Z^{-1/2}F_{\mu \nu}= F^{(P)}_{\mu \nu} \;,\; Z^{-1/2}A_{\mu} = A^{(P)}_{\mu} \;,\; \psi = \psi^{(P)} \;,\; \varphi = \varphi^{(P)}
\end{equation}

\vspace{0.2cm}

\ni and

\vspace{0.2cm}

\begin{equation}  
Z^{1/2} e = e^{(P)} \;,\; Z f = f^{(P)} \;,\; Z^{-1/2}\zeta = \zeta^{(P)} \;,\; M = M^{(P)} \;,\; m_\psi = m_\psi^{(P)} \;,\; m_\varphi = m_\varphi^{(P)}
\end{equation}

\ni (and, in consequence, $A_{\mu \nu} = A^{(P)}_{\mu \nu}$ and $Z^{1/2}j_{\mu} = j^{(P)}_{\mu}$ as $j_\mu \propto e$). In fact, from Eqs. (30), (32) and (33) we infer that the Lagrangian (29) equal to (30) is equal also to the form 

\begin{eqnarray}
{\cal{L}}_{\rm eff} & = &  -\frac{1}{4}F^{(P)}_{\mu \nu} F^{\mu \nu\,(P)}- j^{(P)}_\mu A^{\mu \,(P)} - \frac{1}{4}\frac{f^{(P)}}{M^2} (2\! <\!\!\varphi\!\!>_{\rm vac}\varphi^{(\rm ph)} + \varphi^{(\rm ph)\,2}) F^{(P)}_{\mu \nu} F^{\mu \nu\,(P)} \nonumber \\ & &
- \frac{1}{2}\frac{f^{(P)} \zeta^{(P)}}{M^2} \varphi F^{(P)}_{\mu \nu} (\bar{\psi}\sigma^{\mu \nu}\psi)  - 
\frac{1}{4}\frac{f^{(P)} \zeta^{(P)\,2}}{M^2} (\bar{\psi}\sigma_{\mu \nu}\psi)(\bar{\psi}\sigma^{\mu \nu}\psi) \nonumber \\ & & + (\psi\!{\rm -}\;{\rm and} \;\varphi^{(\rm ph)}\!{\rm -terms\; as\; in \;Eq.\; (21)}) \equiv  {\cal{L}}^{(P)}_{\rm eff}\;, 
\end{eqnarray}

\vspace{0.2cm}

\ni where $\varphi \equiv {<\!\!\varphi\!\!>_{\rm vac}} + \varphi^{(\rm ph)}$. The rhs of Eq. (34) defines the Lagrangian as a function of $P$-renormalized quantities. Further on, their label $P$ can be dropped.

Obviously, the quantities after the $P$-renormalization are still bare quantities, subject to the conventional infinite (higher-order) renormalization (if it can be performed). Notice, that the transformation defined by Eqs. (32) and (33) differs from that discussed in Ref. [1] by changing now the constant $\zeta$ instead of the sterino field $\psi$ as was before.

\vspace{0.3cm}

\ni {\bf 4. Sterinos as the thermal dark matter} 

\vspace{0.3cm}

In our model of hidden sector of the Universe, it is natural to consider sterinos as candidates for the thermal cold dark matter. Then, in the early Universe, sterinos $\psi$ are immersed in the thermal-equilibrium bath consisting of many Standard-Model and a few sterile degrees of freedom (in fact, only $\psi, \varphi $ and $A_{\mu \nu}$). In this case, freeze-out processes lead to a relict abundance of sterinos, providing today's cold dark matter. The abundance of dark matter presently observed by WAMP, $\Omega_{\rm DM}\,  h^2 \simeq 0.1 $, implies that the thermal average of total annihilation cross-section of a sterino-antisterino pair multiplied by sterino relative velocity can be estimated thermodynamically as [4]

\begin{equation}
<\!\sigma_{\rm ann}\, v_{\rm DM}\!> \,\simeq\, {\rm pb} \,\simeq \,\frac{8}{\pi} \frac{10^{-3}}{{\rm TeV}^2}
\end{equation}

\ni (pb = $10^{-36}$ cm$^2$, $c = 1 = \hbar$). This estimation is the same as that for the popular model [4], where cold dark matter consists of weakly interacting massive particles (WIMPs) that, likely, are identical with neutralinos appearing in supersymmetric extensions of~the Standard Model. This provides us with a strong argument for supersymmetry as a real possibility. But, it does not exclude sterinos $\psi$ with an adequate mass as possible candidates for cold dark matter. Then, the leading annihilation channels would be $ \bar\psi \psi \rightarrow \varphi^{\rm (ph)}\gamma$ and $ \bar\psi \psi \rightarrow e^+e^-$ with an adequate strength. Sterons $ \varphi^{\rm (ph)}$, being unstable with the leading decay channels $\varphi^{\rm (ph)}\rightarrow \gamma\gamma$ and $\varphi^{\rm (ph)} \rightarrow e^+e^- \gamma$, cannot be taken into account as cold dark matter, unless their mass is small enough to make steron life-time comparable with the enormous age of the Universe.

With the sterino interaction

\begin{equation}
-\frac{1}{2}\; \frac{f \zeta}{M^2} \varphi^{\rm(ph)} F_{\mu\,\nu}(\bar\psi \sigma^{\mu\,\nu} \psi ) 
\end{equation}

\ni (see Eq. (26)) or with the sterino interaction (27) (resulting also from Eq. (26)) plus the \SM electromagnetic coupling for electrons

\begin{equation}
-\frac{1}{2}\; \frac{f \zeta}{M^2}{<\!\!\varphi\!\!>_{\rm vac}} F_{\mu\,\nu}  (\bar\psi \sigma^{\mu\,\nu} \psi ) + e\,\bar{\psi}_e\,\gamma^\mu \,\psi_e A_\mu
\end{equation}

\ni ($e = |e|$) we can calculate in the sterino-antisterino centre-of-mass frame (where $v_{\rm DM} = 2v_\psi $) the following annihilation cross-sections:

\begin{equation}
\sigma(\bar{\psi} \psi \rightarrow \gamma \varphi^{\rm (ph)}) 2v_\psi = \frac{1}{6\pi} 
\left(\frac{f \zeta}{M^2}\right)^{\!\!2}
\left(E^2_\psi + 2m^2_\psi\right) \left(1 -   \frac{m^2_\varphi}{4 E^2_\psi}\right) 
\end{equation}

\ni or

\begin{equation}
\sigma(\bar{\psi} \psi \rightarrow e^+ e^-) 2v_\psi = \frac{1}{12\pi} 
\left(\frac{e f \zeta<\!\!\varphi\!\!>_{\rm vac}}{M^2}\right)^{\!\!2}\frac{E^2_\psi + 2m^2_\psi}{E^2_\psi} \,,
\end{equation}

\ni respectively, where in the second channel the electron mass is neglected ($E_e = E_\psi \geq m_\psi \gg m_e$). Hence, their ratio is given as

\begin{equation}
\frac{\sigma(\bar{\psi} \psi \rightarrow e^+ e^-)}{\sigma(\bar{\psi} \psi \rightarrow \gamma \varphi^{\rm (ph)})}= \frac{e^2}{2} <\!\!\varphi\!\!>^2_{\rm vac} \left(E^2_\psi - \frac{m^2_\varphi}{4}\right)^{-1} \leq  \frac{e^2}{2} \frac{<\!\!\varphi\!\!>^2_{\rm vac}}{m^2_\psi} \left(1 - \frac{m^2_\varphi}{4m^2_\psi} \right)^{-1}
\end{equation}

\ni since $E_\psi \geq m_\psi $. If, tentatively, $m^2_\psi \sim m^2_\varphi \sim$ (1 to $10^{-3}) <\!\!\varphi\!\!>^2_{\rm vac}$ ({\it i.e.}, $\xi^2 \sim 2\lambda \sim$ (1 to $10^{-3}$), see Eq. (13)), then 

\begin{equation}
\frac{\sigma(\bar{\psi} \psi \rightarrow e^+ e^-)}{\sigma(\bar{\psi} \psi \rightarrow \gamma \varphi^{\rm (ph)})}\stackrel{<}{\sim} \frac{2}{3}\, e^2\, {\rm(1\; to}\; 10^{3}) = 0.0611 \;{\rm to }\; 61.1 
\end{equation}

\ni ($e^2 = 4\pi \alpha = 0.0917$ with $\alpha = 1/137$) and so, approximately

\begin{equation}
\sigma_{\rm ann} v_{\rm {DM}} \simeq [\sigma(\bar{\psi} \psi \rightarrow \gamma \varphi^{\rm (ph)}) + \sigma(\bar{\psi} \psi \rightarrow e^+ e^-)] 2v_\psi \sim \frac{3}{8\pi} \left(\frac{f \zeta}{M^2}\right)^{\!2} m^2_\psi\;(1.06\;{\rm to}\;62.1) \,,
\end{equation}

\ni when $E_\psi \simeq m_\psi$ ({\it i.e.}, $\vec{p}^{\;2}_\psi/m^2_\psi\ll 1$). In this approximation, $<\!\!\sigma_{\rm ann} v_{\rm DM}\!\!>\, \simeq \sigma_{\rm ann} v_{\rm {DM}}$.

By comparing Eqs. (35) and (42) we can estimate the sterino mass: 

\begin{equation}
m_\psi\! \sim \!{\rm(1\, to}\, 10^{-3})\frac{M^2}{m_\psi} \sim {\rm(1\, to}\, 10^{-3})\frac{\sqrt{3}}{8}f \zeta \,(\sqrt{1.06}\,{\rm to}\,\sqrt{62.1})\times 10^{3/2} \,{\rm TeV}  = {\rm(650\, to}\, 4.9) \zeta\,{\rm GeV} \,,
\end{equation}

\ni if, tentatively, in addition $m^2_\psi \sim {\rm(1\; to}\; 10^{-3}) M^2$ ({\it i.e.}, $\xi^2 \sim {\rm(1\; to}\; 
10^{-3}) \eta/2$, see Eqs. (13) and (15)), so $M^2 \sim <\!\!\varphi\!\!>_{\rm vac}^2$ ({\it i.e.}, $\eta/2 \sim 1$, see Eq. (15)), and we decide to make a bold conjecture that $f = e^2 = 0.0917.$ In Eq. (43) it is natural to expect that $\zeta \sim 1$. From Eq. (43) we find the following estimation for the mass of $A$ bosons: $ M \sim (1\;{\rm to}\;10^{3/2}) m_\psi \sim (650\;{\rm to}\;160)\zeta $ GeV. Note that the stronger is the dominance of $M^2$ over $m^2_\psi \sim m^2_\varphi $ (Eq. (43)), the easier the dominance of  $M^2$ over $-\Box$ in virtual states in Eq. (23) (at any rate for nonrelativistic processes).

Note that for $f = e^2$ our supplemented Maxwell equations (1) and (2) take the neat form

\begin{equation}
\partial_\nu (F^{\mu\,\nu} + e \,\varphi\, A^{\mu\,\nu}) = -j^\mu 
\end{equation}

\ni and

\begin{equation}
(\Box - M^2)A_{\mu\,\nu} = - e\, (\varphi F_{\mu\,\nu} + \zeta \,\bar\psi \,\sigma_{\mu\,\nu} \,\psi)\;,  
\end{equation}

\ni where $F_{\mu\,\nu} = \partial_\mu A_\nu - \partial_\nu A_\mu$. In this case, the sterino magnetic moment generated spontaneously by $<\!\!\varphi\!\!>_{\rm vac} \neq 0$ becomes (see Eq. (28) with ${\rm(1\; to}\; 
10^{3/2})\, \xi \sim \sqrt{\eta/2} \sim 1$):  

\begin{equation}
\mu_\psi \sim {\rm(1\; to}\; 10^{-3/2}) \frac{e^2 \zeta}{2m_\psi} \sim (0.071 \;{\rm to}\; 0.29)\;{\rm TeV}^{-1}
\end{equation}

\ni with $m_\psi \sim (0.65\;{\rm to}\; 0.0049)\,\zeta \;{\rm TeV}$ (in Eq. (46) the factor $\zeta$ is cancelled). 

In contrast to sterinos, sterons are unstable. Their simplest decay channel $\varphi^{\rm (ph)}\!\! \rightarrow \!\!
\gamma\gamma$ gets the following decay rate at rest implied by the steron interaction $-(1/4)(f/M^2)\, \varphi\, F_{\mu \nu}\,\varphi\, F^{\mu \nu}$ (see Eq. (26)):

\begin{eqnarray}
\Gamma(\varphi^{(\rm ph)}\!\rightarrow\! \gamma\gamma) & = & \frac{1}{128\pi} \!\left(\frac{f<\!\!\varphi\!\!>_{\!\rm vac}}{M^2} \right)^{\!\!2} \,m^3_\varphi  \sim {\rm(1\; to}\; 10^{-3})\,\frac{e^4}{128\pi} m_\varphi  \nonumber \\ & \sim & (14\;{\rm to}\;1.0\times 10^{-4}) \,\zeta\,{\rm MeV} \;, 
\end{eqnarray}

\ni if, tentatively, ${\rm(1\; to}\; 10^{3})\, m^2_\psi \sim {\rm(1\; to}\; 10^{3})\, m^2_\varphi \sim M^2 \sim \,<\!\!\varphi\!\!>_{\!\rm vac}^2$ ({\it i.e.}, ${\rm(1\; to}\; 10^{3})\,\xi^2 \sim {\rm(1\; to}\; 10^{3})\,2\lambda \sim \eta/2 \sim 1$, see Eqs. (13) and (15)). In Eq. (47) it is natural to expect that $\zeta \sim 1$.

Also sterile mediating bosons, quanta of the field $A_{\mu \nu}$ or {\it A} bosons, are unstable, since due to the interaction

$$
-\frac{1}{2}\; \sqrt{f}{<\!\!\varphi\!\!>_{\rm vac}}F_{\mu\,\nu} A^{\mu\,\nu} + e \bar\psi_e \gamma^{\mu} \psi_e  A_\mu \eqno{(48{\rm a})}
$$

\ni (see Eq. (16)) they can decay through virtual photons into electron-positron pairs: $A \rightarrow \gamma^* \rightarrow e^+ e^-$ ($M \gg 2m_e$). The production of $A$ bosons can be caused by the interaction

$$
-\frac{1}{2}\; \sqrt{f} \varphi^{\rm (ph)} F_{\mu\,\nu} A^{\mu\,\nu} + e \bar\psi_e \gamma^{\mu} \psi_e  A_\mu \eqno{(48{\rm b})}
$$

\ni (see Eq. (16)) leading to the annihilation of electron-positron pairs into physical sterons and $A$ bosons: $e^+ e^- \rightarrow \gamma^* \rightarrow \varphi^{\rm (ph)}A$. Here, the centre-of-mass threshold energy is 
$m_\varphi + M \sim {(1.3\;\rm to\;0.16)\;\zeta TeV}$, if our tentative estimations of $m_\psi \sim m_\varphi$ and $M$ are applied. Analogically, one may consider the production process $\bar{p} p \rightarrow \gamma^* \rightarrow \varphi^{\rm (ph)} A$. By means of the first term of interaction (48b) the direct decay $A \rightarrow \varphi^{\rm (ph)} \gamma$ is allowed if $M > m_\varphi$ (even if $M \sim m_\varphi$), otherwise it is forbidden.

Sterinos, due to their magnetic interaction (27), can be produced in sterino-antisterino pairs through virtual photons in high-energy collisions of \SM charged particles. The simplest process of this kind is $e^+ e^- \rightarrow \gamma^* \rightarrow \bar\psi \psi$, inverse to the annihilation channel $\bar\psi \psi \rightarrow \gamma^* \rightarrow e^+ e^-$. In this case, the centre-of-mass threshold energy is $2m_\psi \sim (1.3\;{\rm to\;0.0098) \zeta\, TeV}$ due to our tentative estimation of $m_\psi$ With the use of interaction (37) we obtain in the electron-positron centre-of-mass frame the following cross-section:

\addtocounter{equation}{+1}

\begin{eqnarray}
\sigma(e^+e^- \!\rightarrow \bar{\psi} \psi) 2v_e & = & \frac{1}{12\pi}\!  \left(\frac{e f \zeta<\!\!\varphi\!\!>_{\rm vac}}{M^2}\right)^{\!\!2}\!\!\left(1+\frac{m^2_\psi}{2E^2_e}\right) \sqrt{1-\frac{m^2_\psi}{E^2_e}} \nonumber \\ & \sim & {\rm(1\; to}\; 10^{-3})\,\frac{1}{12\pi}  \left(\frac{e^3 \zeta}{m_\psi}\right)^{\!\!2}\!\!\left(1+\frac{m^2_\psi}{2E^2_e}\right) \sqrt{1-\frac{m^2_\psi}{E^2_e}} \,,
\end{eqnarray}

\ni where the electron mass is neglected ($E_e = E_\psi \geq m_\psi \gg m_e$). In the second step of Eq. (49), we use our tentative assumptions $ {\rm(1\; to}\, 10^{3})\,m_\psi^2 \sim M^2 \sim <\!\!\varphi\!\!>_{\rm vac}^2$ ({\it i.e.}, ${\rm(1\; to}\, 10^{3}) \xi^2 \sim \eta/2 \sim 1$, see Eqs. (13) and (15)) and $f = e^2 =0.0917$. In the cross-section (49), its numerical coefficient can be estimated as

\begin{equation}
\frac{{\rm(1\; to}\; 10^{-3})\,e^6 \,\zeta^2 }{12\pi\, m^2_\psi} \sim {\rm(4.9\; to}\; 83)\times 10^{-5} \;{\rm TeV}^{-2}
\end{equation}

\ni with $m_\psi \sim $ (0.65 to 0.0049) $\zeta$ TeV (here, the factor $\zeta^2 $ is cancelled). A formula analogical to Eq. (49) holds also for the process  $\bar{p} p \rightarrow \gamma^* \rightarrow \bar\psi \psi$, if protons are treated as point-like particles.

The unstable $A$ bosons can be produced in pairs $AA$ through virtual photon pairs $\gamma^*\gamma^*$. The simplest process of such a production is $e^+ e^- \rightarrow \gamma^*\gamma^* \rightarrow AA$ (where $\gamma^*\rightarrow A$). The simplest production process of all for the pair $AA$ is the sterino-antisterino annihilation channel $\bar\psi \psi \rightarrow \bar\psi^* \psi^* AA \rightarrow AA$ (where $ \bar\psi^* \psi^* \rightarrow $ vac).

Concluding, if in our model of hidden sector with the photonic portal we put tentatively $m^2_\psi\! \sim\! m^2_\varphi\! \sim {\rm(1\; to}\; 10^{-3}) \!M^2  \sim {\rm(1\; to}\; 10^{-3})<\!\!\varphi\!\!>_{\!\rm vac}^2\!$ ({\it i.e.}, $\xi^2 \!\sim\! 2\lambda \!\sim\! {\rm(1\; to}\; 10^{-3})\eta/2 \sim {\rm(1\; to}\; 10^{-3})$) and $f = e^2$, then for  ${m_\psi \sim\,{\rm(0.6\; to}\;0.005)\zeta\,{\rm TeV}}$ our sterinos can be candidates for thermal cold dark matter, whereas our sterons are unstable with the decay rate $\sim {\rm(10\; to}\; 10^{-4}) \zeta$ MeV (it is natural to expect that $\zeta \sim 1$). It is interesting enough to realize that sterinos get tiny magnetic moment $\mu_\psi \sim (0.07\;{\rm to}\;0.3)$ TeV$^{-1}$ spontaneously generated by $ <\!\!\varphi\!\!>_{\!\rm vac} \neq 0$ and so, can be polarized in external magnetic fields. The same nonzero vacuum expectation value $ <\!\!\varphi\!\!>_{\!\rm vac} \neq 0$ of the steron field generates spontaneously three masses $m_\psi, m_\varphi$ and $M$. Due to sterino magnetic interaction, the sterino-antisterino pairs can be produced through virtual photons emitted in 
high-energy collisions of \SM charged particles.

Recently, the DAMA collaboration have reinforced their claim to have found evidence for dark-matter particles with a relatively small mass in the GeV range [5]. Taking into account the limits imposed by null results of other direct-detection experiments for dark matter, the DAMA results may correspond to WIMPs with the mass 3 to 8 GeV (so, in our model, to sterinos with such a moderate mass, say, $m_\psi \sim 5$ GeV). 
 
\vfill\eject

\vspace{0.4cm}

{\centerline{\bf References}}

\vspace{0.4cm}

\baselineskip 0.73cm

{\everypar={\hangindent=0.65truecm}
\parindent=0pt\frenchspacing

{\everypar={\hangindent=0.65truecm}
\parindent=0pt\frenchspacing

~[1]~W.~Kr\'{o}likowski,  {\it Act. Phys. Polon.} {\bf B 39}, 1881 (2008) (an extended version of arXiv: 0712.0505 [{\tt hep--ph}]); arXiv: 0803.2977 [{\tt hep--ph}]; arXiv: 0806.2036 [{\tt hep--ph}].

\vspace{0.2cm}

~[2]~For recent publications {\it cf.} J. March-Russell, S.M. West, D. Cumberbath and D.~Hooper, {\it JHEP} {\bf 0807}, 058 (2008); D.~Hooper and K.M. Zurek,  {\it Phys. Rev.} {\bf D 77}, 087302 (2008); J.~McDonald and 
N.~Sahu, {\it JCAP} {\bf 0806}, 026 (2008); Y.G.~Kim, K.Y.~Lee and S.~Shin, {\it JHEP} {\bf 0805}, 100 (2008); J.L.~Feng and J.~Kumar, arXiv: 0803.4196 [{\tt hep-ph}]; R.~Foot, arXiv: 0804.4518 [{\tt hep-ph}]; and references therein.

\vspace{0.2cm}

~[3]~Particle Data Group, {\it Review of Particle Physics, J. Phys}, {\bf G 33}, 1 (2006).

\vspace{0.2cm}

~[4]~For recent reviews {\it cf.} G. Bartone, D.~Hooper and J.~Silk, {\it Phys. Rept.} {\bf 405}, 279 (2005); M.~Taoso, G.~Bartone and A.~Masiero, arXiv: 0711.4996 [{\tt astro-ph}]; {\it cf.} also E.W.~Kolb and S.~Turner, {\it Early Universe} (Addison-Wesley, Reading, Mass., 1994); K.~Griest and D.~Seckel, {\it Phys. Rev.} {\bf D 43}, 3191 (1991);J.L.~Fang, H.~Tu and H.-B.~Yu, arXiv: 0808.2318 [{\tt hep-ph}].  

\vspace{0.2cm}

~[5]~R. Bernabei {\it et al}. (DAMA Collaboration), , arXiv: 0804.2741 [{\tt astro-ph}]; for a discussion {\it cf.} 
D.~Hooper, F.~Petriello, K.M.~Zurek and M.~Kamionkowski, arXiv: 0808.2464 [{\tt hep-ph}].

\vfill\eject

\end{document}